\documentclass[preprint,showpacs,preprintnumbers,amsmath,amssymb]{revtex4}

\usepackage{graphicx}
\usepackage{dcolumn}
\usepackage{bm}

\begin{document}

\preprint{Submission to Phys. Rev. B}

\title{
Spiral magnetic structure in spin-5/2 frustrated trimerized chains in SrMn$_3$P$_4$O$_{14}$
}

\author{Masashi Hase$^{1}$}
 \email{HASE.Masashi@nims.go.jp}
\author{Vladimir Yu. Pomjakushin$^2$}
\author{Lukas Keller$^2$}
\author{Andreas D\"onni$^1$}
\author{Osamu Sakai$^1$}
\author{Tao Yang$^3$}
\author{Rihong Cong$^3$}
\author{Jianhua Lin$^3$}
\author{Kiyoshi Ozawa$^1$}
\author{Hideaki Kitazawa$^1$}

\affiliation{%
${}^{1}$National Institute for Materials Science (NIMS), 1-2-1 Sengen, 
Tsukuba, Ibaraki 305-0047, Japan \\
${}^{2}$Laboratory for Neutron Scattering, Paul Scherrer Institut (PSI), 
CH-5232 Villigen PSI, Switzerland \\
${}^{3}$College of Chemistry and Molecular Engineering, 
Peking University, Beijing 100871, People's Republic of China
}%

\date{\today}

\begin{abstract}

We study a spin-5/2 antiferromagnetic trimerized chain substance SrMn$_3$P$_4$O$_{14}$ 
using neutron powder diffraction experiments. 
The coplanar spiral magnetic structure appears below $T_{\rm N1} = 2.2(1)$ K.  
Values of several magnetic structure parameters change rapidly at $T_{\rm N2} = 1.75(5)$ K, 
indicating another phase transition, 
although the magnetic structures above and below $T_{\rm N2}$ are the qualitatively same.  
The spiral magnetic structure can be explained by 
frustration between nearest-neighbor and next-nearest-neighbor exchange interactions 
in the trimerized chains. 

\end{abstract}


\maketitle

\section{Introduction}

Frustrated magnets can exhibit intriguing magnetic states such as 
the quantum spin-liquid state,\cite{Anderson73} the chiral ordered state,\cite{Miyashita84,Onoda07} 
the spin nematic or the multipolar state,\cite{Tsunetsugu06,Zhitomirsky08} and the spin-gel state.\cite{Kawamura10} 
Therefore, frustrated magnets have been investigated extensively. 
Among various frustrated magnets, frustrated spin chains provide ideal grounds for studies.\cite{Haldane82a,Haldane82b}
Frustrated spin chains are simple models that still show surprisingly rich physics. 
For example, in the frustrated spin-1/2 chain with ferromagnetic nearest-neighbor and 
antiferromagnetic next-nearest-neighbor exchange interactions ($J_{\bf NN}$ and $J_{\bf NNN}$ interactions, respectively) 
in the presence of magnetic fields, 
theoretical studies predict various ground-state phases including the vector chiral phase, the nematic phase, other multipolar phases, and the spin-density-wave 
phases.\cite{Chubukov91,Kolezhuk05,Heidrich06,Vekua07,Kecke07,Hikihara08,Sudan09,Sato09,Heidrich09,Sato11} 
Several model substances have been found and are summarized in Table 1 in Ref. [\onlinecite{Hase04}] or Fig. 5 in Ref. [\onlinecite{Drechsler07}]. 
Experimental results are compared with theoretical results.\cite{Hikihara08,Sato11}

Theoretical investigations were performed in frustrated longer-period spin chains 
such as alternating chains,\cite{Brehmer96,Nakamura97,Brehmer98,Watanabe99} 
while there is almost no model substance. 
To our knowledge, the only example is CuGeO$_3$. 
The frustrated alternating spin-1/2 chain is realized in the spin-Peierls state.\cite{Hase93a,Hase93b,Hase93c,Hase95,Castilla95,Riera95,Hase96,Martin97} 
In most of one-dimensional magnets, the spin-spin distance in the $J_{\bf NNN}$ interaction is large. 
Therefore, the $J_{\bf NNN}$ interaction is usually much smaller than the $J_{\bf NN}$ interaction.  
It is difficult to find model substances having frustrated spin chains. 

We have paid our attention to SrMn$_3$P$_4$O$_{14}$. 
We expected the $J_1 - J_1 - J_2$ trimerized chain shown in Fig. 1.\cite{Yang08} 
The magnetization results could be explained using 
the spin-5/2 trimer formed by the antiferromagnetic (AF) $J_1$ interaction ($J_1 /k = 4.0$ K).\cite{Hase09} 
Magnetic excitations were observed in inelastic neutron scattering experiments and 
could be also explained using the trimer model.\cite{Hase11}  
Therefore, the $J_2$ value is small in comparison with the $J_1$ value. 
A magnetic phase transition appears at about 2.6 K.\cite{Yang08} 
Therefore, the $J_2$ interaction is not negligible. 
The $J_2$ interaction may be comparable to NNN exchange interactions in the chain. 
We expected the occurrence of an unconventional magnetic structure and performed neutron powder diffraction experiments.
In this paper, we report that SrMn$_3$P$_4$O$_{14}$ shows a coplanar spiral magnetic structure. 
In the classical limit, the ground state of the frustrated uniform spin chain has a spiral magnetic structure 
with a pitch angle $\cos^{-1} (- J_{\bf NN} /4 J_{\bf NNN})$ for $|J_{\bf NN} / J_{\bf NNN}| < 4$.\cite{Bursil95} 
Similarly, we can explain the spiral magnetic structure in SrMn$_3$P$_4$O$_{14}$
taking the $J_1$,  $J_2$, and NNN exchange interactions in the chain into account. 

\begin{figure}
\begin{center}
\includegraphics[width=8cm]{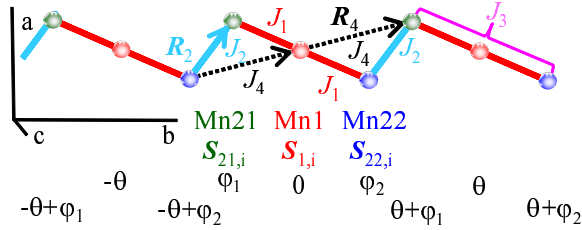}
\caption{
(Color online)
The spin system in SrMn$_3$P$_4$O$_{14}$. 
Mn$^{2+}$ ions ($3d^5$) have localized spin 5/2. 
Two kinds of short Mn-Mn bonds exist. 
The Mn-Mn distances are 3.26 and 3.31 \AA \ at 4.0 K.
The exchange interaction parameters are respectively defined as $J_1$ and $J_2$. 
The $J_1$ and $J_2$ interactions form a trimerized chain that is almost parallel to the $b$ axis. 
The value of the antiferromagnetic $J_1$ interaction was evaluated as 3.4 to 4.0 K.\cite{Hase09,Hase11}
The $J_3$ and $J_4$ interactions are the next-nearest-neighbor exchange interactions in the chain. 
${\bf S}_{1,i}$, ${\bf S}_{21,i}$, and ${\bf S}_{22,i}$ are spin operators. 
We use the parameters $\theta$, $\varphi_1$, and $\varphi_2$ 
to show angles between magnetic moments in a coplanar spiral magnetic structure. 
${\bf R}_2$ and ${\bf R}_4$ indicate the positional vectors in 
the $J_2$ and $J_4$ Mn-Mn bonds, respectively. 
}
\end{center}
\end{figure}

\section{Experiment}

We synthesized SrMn$_3$P$_4$O$_{14}$ powders under hydrothermal conditions.\cite{Yang08}
We carried out neutron powder diffraction experiments 
at the Swiss spallation neutron source SINQ in Paul Scherrer Institute.  
We used the high-resolution powder diffractometer for thermal neutrons HRPT 
(wavelength $\lambda=1.886$~\AA, high intensity mode $\Delta d/d\geq1.8\times 10^{-3}$) \cite{hrpt} and 
the high-intensity cold neutron powder diffractometer (DMC) with $\lambda = 2.458$~\AA. 
We performed Rietveld refinements of crystal and magnetic structures
using the {\tt FULLPROF Suite}  program package~\cite{Rodriguez93}  
with the use of its internal tables for scattering lengths and magnetic form factors. 

\section{Results}

Figure 2 depicts the neutron powder diffraction pattern of SrMn$_3$P$_4$O$_{14}$ 
recorded using the HRPT diffractometer with $\lambda = 1.886$ \AA \ at 4.0 K (paramagnetic state).  
The crystal structure model in the space group $P2_1 /c$ (No. 14) proposed in [\onlinecite{ Yang08}] fits well our data. 
The refined structure parameters are presented in Table I. 
They were kept fixed in the subsequent refinements of the magnetic structure. 

\begin{figure}
\begin{center}
\includegraphics[width=8cm]{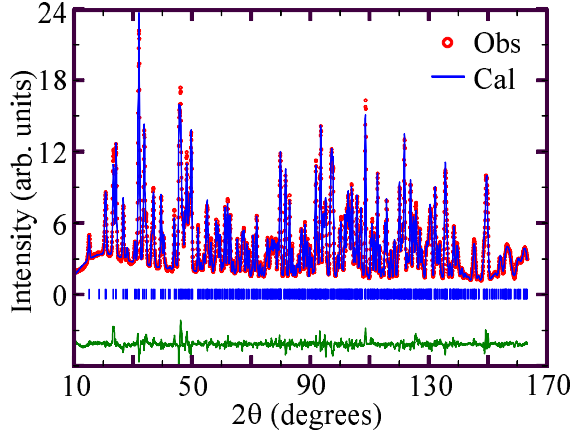}
\caption{
(Color online)
Neutron powder diffraction pattern of SrMn$_3$P$_4$O$_{14}$ 
at 4.0 K measured using the HRPT diffractometer ($\lambda = 1.886$ \AA). 
Lines on the observed pattern and at the bottom 
show a Rietveld refined pattern and the difference between the observed and 
the Rietveld refined patterns. 
Hash marks represent the positions of nuclear reflections.
}
\end{center}
\end{figure}

\begin{table*}
\caption{\label{table1}
Structural parameters of SrMn$_3$P$_4$O$_{14}$ derived from Rietveld refinement of
the HRPT neutron powder diffraction pattern at 4.0 K.
The space group is monoclinic $P2_1 /c$ (No. 14).  
The lattice constants at 4.0 K are
$a = 7.661(1)$ \AA, $b = 7.784(1)$ \AA, $c = 9.638(1)$ \AA , and $\beta = 111.70(2)^{\circ}$.
Estimated standard deviations are shown in parentheses. 
The atomic displacement parameters $B_{\rm iso}$ were constrained to be the same for 
Mn and P. 
The reliability factors of the refinement are  
$R_{\rm wp}=5.11$\%, $R_{\rm exp}=1.53$\%, $\chi^2=11.2$, and $R_{\rm Bragg}=4.04$\%.
}
\begin{ruledtabular}
\begin{tabular}{cclllr}
Atom & Site & $x$ & $y$ & $z$ & $B_{\rm iso}$ \AA$^2$  \\
\hline
Sr & 2{\it a} & 0 & 0 & 0 & 0.09(5) \\
Mn1 & 2{\it b} & 0.5 & 0 & 0 & 0.11(5) \\
Mn2 & 4{\it e} & 0.6895(5) & 0.1193(5) & 0.5332(4) & 0.11(5) \\ 
P1 & 4{\it e} & 0.6186(4) & 0.7051(4) & 0.8061(3) & 0.23(4) \\ 
P2 & 4{\it e} & 0.8937(4) & 0.0600(3) & 0.2966(3) & 0.23(4) \\
 O1 & 4{\it e} & 0.1228(3) & 0.6832(3) & 0.0903(3) & 0.07(4) \\       
 O2 & 4{\it e} & 0.3326(3) & 0.1255(3) & 0.1123(3) & 0.15(4) \\       
 O3 & 4{\it e} & 0.8187(3) & 0.8764(3) & 0.3219(3) & 0.12(4) \\       
 O4 & 4{\it e} & 0.4650(3) & 0.0820(3) & 0.6128(2) & 0.03(4) \\       
 O5 & 4{\it e} & 0.7694(3) & 0.1162(3) & 0.1351(2) & 0.22(4) \\       
 O6 & 4{\it e} & 0.5100(3) & 0.7143(3) & 0.6411(3) & -0.02(4) \\  
 O7 & 4{\it e} & 0.0910(3) & 0.0213(3) & 0.2991(3) & 0.17(4) \\        
\end{tabular}
\end{ruledtabular}
\end{table*}

Figure 3 depicts fragments of diffraction patterns 
recorded using the DMC diffractometer. 
We observed many new reflections below $T_{\rm N1} = 2.2(1)$ K. 
One of them is shown in Fig. 3(a).  
The magnetic susceptibility data indicate a magnetic phase transition at low $T$.\cite{Yang08} 
Therefore, the new reflections are magnetic reflections.  
The position of the reflection in Fig. 3(a) is shifted clearly and the intensity increases rapidly between 1.7 and 1.8 K, 
indicating another phase transition.  
Diffuse scattering is readily apparent in Fig. 3(b). 
Several magnetic reflections exist between 20 and 35$^{\circ}$ below $T_{\rm N1}$. 
Therefore, the diffuse scattering stems from magnetic correlations. 
The shape of the diffuse scattering resembles one- or two-dimensional magnetic Bragg scattering 
with a cutoff at low Q and long tail at high Q (scattering vector magnitude).  
As the temperature $T$ is lowered, the intensity around 25$^{\circ}$ increases down to 2.6 K and decreases below 2.6 K. 
The diffuse scattering remains even at 1.5 K. 

\begin{figure}
\begin{center}
\includegraphics[width=8cm]{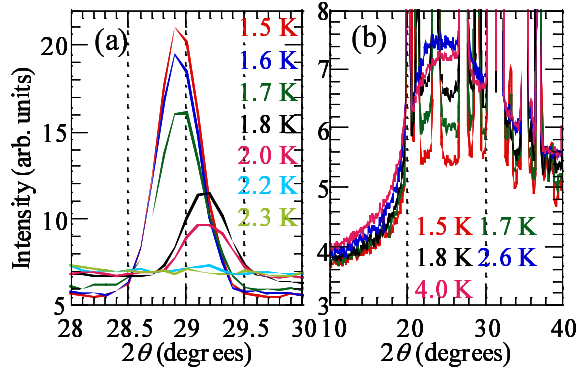}
\caption{
(Color online)
Neutron powder diffraction patterns of SrMn$_3$P$_4$O$_{14}$ 
measured using the DMC diffractometer ($\lambda = 2.458$ \AA). 
The temperature $T$ of each curve is understood from the following $T$ dependence. 
(a)
The peak intensity decreases monotonically with increasing $T$. 
(b)
The intensity at 25$^{\circ}$ is smallest at 1.5 K. 
As $T$ is raised, the intensity at 25$^{\circ}$ increases and 
is largest at 2.6 K. 
The pattern at 4.0 K is positioned just below the 2.6 K pattern. 
}
\end{center}
\end{figure}

Circles in Fig. 4 represent the neutron powder diffraction pattern 
at 1.5 K recorded using the DMC diffractometer. 
The magnetic reflections can all be indexed 
with a propagation vector {\bf k} = [0, $k_y$, 0] with $k_y \sim 0.32$. 
Using the determined propagation vector we performed the symmetry analysis 
according to Izyumov {\it et al.} \cite{Izyumov91} to derive possible magnetic configurations 
for Mn1 (2d) and Mn2 (4e) sites. 
For the calculation of the basis functions 
we used program $\tt BASIREP$.\cite{Rodriguez93} 
The little group of a propagation vector for the space group $P2_1/c$ contains 
only two symmetry operators: $1$ and $\{2_y|0{1\over2}{1\over2}\}$ 
resulting in the splitting of Mn2 (4e) into two independent orbits Mn21 and Mn22 sites. 
The little group has two 1D irreducible representations (IRs) $\tau_1$ and $\tau_2$ 
with the characters for $2_y$: $\pm e^{-i\pi k_y}$, respectively.   
Both IRs enter three times the magnetic representation for all three Mn-sites. 
The basis functions for $\tau_2$  dictate the absence of the ($0, 1-k, 0$) magnetic reflection, 
which indeed has zero intensity. 
Therefore, we can immediately choose $\tau_2$. 
Nevertheless, we have performed a simulated annealing search of the magnetic structures for both IRs. 
The $\tau_2$ produced excellent results, although the $\tau_1$ failed. 
We found that the $y$-components of magnetic moments can be ignored and  
performed final refinements under the following conditions. 
Only the $x$ and $z$ components of magnetic moments are considered. 
The Mn21 and Mn22 sites are crystallographically identical. 
Therefore, the magnitudes of Mn21 and Mn22 moments are constrained as identical. 
In general, the magnetic structure consists of ellipsoidal spirals running on all three Mn-sites 
with independent phases. 
The spin 5/2 of Mn$^{2+}$ ions is almost isotropic.\cite{Hase09} 
Therefore, we assumed a circular spiral magnetic structure. 

\begin{figure}
\begin{center}
\includegraphics[width=8cm]{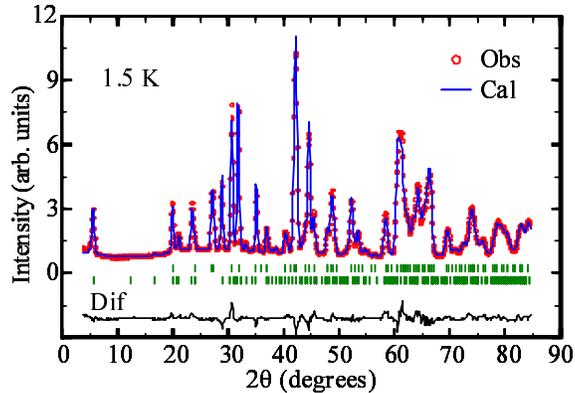}
\caption{
(Color online)
Neutron powder diffraction pattern of SrMn$_3$P$_4$O$_{14}$ at 1.5 K 
measured using the DMC diffractometer ($\lambda = 2.458$ \AA). 
Lines on the observed pattern and at the bottom 
show a Rietveld refined pattern and the difference between the observed and 
the Rietveld refined patterns, respectively. 
Upper and lower Hash marks represent the positions of nuclear and magnetic reflections, respectively. 
}
\end{center}
\end{figure}

The line on the observed pattern at 1.5 K in Fig. 4 shows 
a Rietveld refined pattern including both nuclear and magnetic contributions.  
It can well reproduce the observed pattern. 
Figure 5 depicts the coplanar spiral magnetic structure at 1.5 K. 
The value of $k_y$ is 0.317, indicating an incommensurate magnetic structure. 
Magnetic moments have $x$ and $z$ components. 
The angle between Mn1 and Mn21 (Mn22) moments in each trimer is 166$^{\circ}$ (164$^{\circ}$). 
The angle between Mn21 and Mn22 moments in the $J_2$ bond is 111$^{\circ}$. 
These angles indicate that both the $J_1$ and $J_2$ interactions are AF. 

\begin{figure}
\begin{center}
\includegraphics[width=8cm]{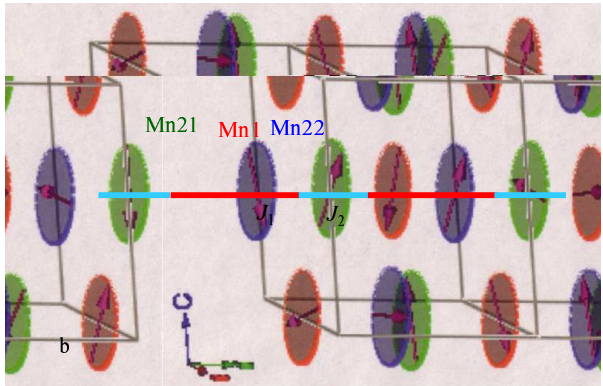}
\caption{
(Color online)
The coplanar spiral magnetic structure of SrMn$_3$P$_4$O$_{14}$ at 1.5 K.  
}
\end{center}
\end{figure}

Figure 6 shows the $T$ dependence of several magnetic structure parameters. 
The value of $k_y$ is abruptly changed between 1.7 and 1.8 K, 
indicating another phase transition at $T_{\rm N2} = 1.75(5)$ K. 
As $T$ is lowered, 
the magnitudes of the Mn1 and Mn2 moments ($M1$ and $M2$) increase. 
The size of magnetic domains $L$ was determined from Lorentzian broadening of the magnetic reflections. 
The value of $L$ increases concomitantly with decreasing $T$. 
Figure 6(c) shows the integrated intensity $I_{\rm ds}$ between $2 \theta = 24.4$ and 26.7$^{\circ}$, 
where diffuse scattering is apparent, as presented in Fig. 3(b). 
As $T$ is lowered below $T_{\rm N1}$, the value of $I_{\rm ds}$ decreases, 
indicating that the diffuse scattering is weakened. 
We emphasize that $M1$, $M2$, $L$, and $I_{\rm ds}$ change rapidly around $T_{\rm N2} = 1.75(5)$ K. 

\begin{figure}
\begin{center}
\includegraphics[width=8cm]{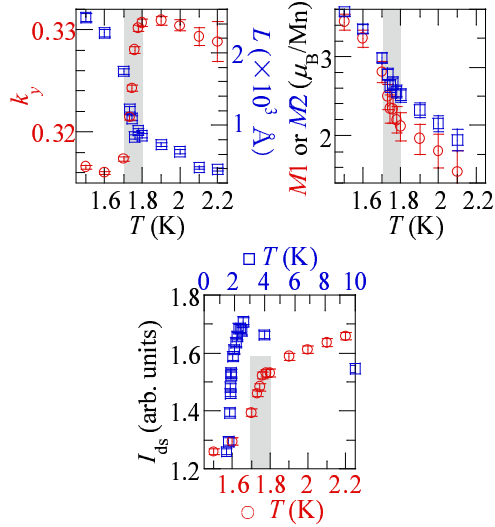}
\caption{
(Color online)
Temperature dependence of several magnetic structure parameters of SrMn$_3$P$_4$O$_{14}$. 
Circles and squares in (a) show the $y$ component of the propagation vector 
(the left vertical axis) and the size of magnetic domains (right), respectively. 
Circles and squares in (b) show the magnitudes of the magnetic moments of Mn1 and Mn2, respectively. 
Integrated intensities of the diffuse magnetic scattering (plus backgrounds) between 24.4 and 26.7$^{\circ}$ 
are shown as circles (lower horizontal scale) and squares (upper) in (c). 
The hatched area shows the transition region around $T_{\rm N2}$. 
}
\end{center}
\end{figure}

\section{Discussion}

We discuss the origin of the spiral magnetic structure in SrMn$_3$P$_4$O$_{14}$. 
Most of spiral magnetic structures are caused by magnetic frustration among plural symmetric exchange interactions. 
As was described in the Section 1, 
magnetic frustration between NN and NNN exchange interactions in a uniform spin chain 
can generate a spiral magnetic structure.\cite{Bursil95} 
We consider whether magnetic frustration between NN and NNN exchange interactions in a trimerized spin chain 
can generate a spiral magnetic structure or not. 
The two kinds of NNN exchange interactions ($J_3$ and $J_4$ interactions) may exist (Fig. 1). 
The Hamiltonian of the exchange interactions is given as follows. 
\begin{eqnarray}
{\cal H}_{\rm ex} &=& \Sigma_i \{
J_1 ({\bf S}_{1,i} \cdot {\bf S}_{21,i} + {\bf S}_{1,i} \cdot {\bf S}_{22,i})  \nonumber \\
& & + J_2 {\bf S}_{21,i} \cdot {\bf S}_{22,i-1} \nonumber \\
& & + J_3 {\bf S}_{21,i} \cdot {\bf S}_{22,i}  \nonumber \\
& & + J_4 ({\bf S}_{1,i} \cdot {\bf S}_{21,i+1} + {\bf S}_{1,i} \cdot {\bf S}_{22,i-1}) \}.
\end{eqnarray}
According to the experimental results, 
we restrict classical ground states of this Hamiltonian to 
states in which magnetic moments lie in the $ac$ plane. 
The relative angle of each magnetic moment from the central Mn1 moment in Fig. 1 
can be defined by one parameter such as $-\theta + \varphi_1$, $-\theta$, and so on. 
The classical exchange energy per trimer is expressed as follows. 
\begin{eqnarray}
E_{\rm ex} &=& (g \mu_{\rm B})^{-2} [ J_1 M_1 M_2 ( \cos \varphi_1 + \cos \varphi_2) \nonumber \\
   & & + J_2 M_2 M_2 \cos (\theta + \varphi_1 - \varphi_2) \nonumber \\
   & & + J_3 M_2 M_2  \cos (\varphi_1 - \varphi_2) \nonumber \\
   & & + J_4 M_1 M_2 \{ \cos (\theta + \varphi_1) + \cos (\theta - \varphi_2) \} ].
\end{eqnarray}
We examine whether we can reproduce the experimental values of $\theta$, 
114$^{\circ}$ and 119$^{\circ}$ at 1.5 K and 1.8 K, respectively,  
under the condition that the experimental values of $\varphi_1$, $\varphi_2$, and $M_1 / M_2 \equiv \alpha$ are given. 
The experimental values of $\varphi_1$, $\varphi_2$, and $\alpha$ are respectively 
166$^{\circ}$, 164$^{\circ}$, and 0.96 at 1.5 K, and 
170$^{\circ}$, 160$^{\circ}$, and 0.84 at 1.8 K. 
The $\theta$ dependent part of the classical exchange energy $E'( \theta )$ is given in Eq. (3). 
\begin{eqnarray}
\frac{E'( \theta )_{\rm ex}  (g \mu_{\rm B})^2}{J_2 M_2^2}
= \cos ( \theta + \varphi_1 - \varphi_2 ) \nonumber \\
& & + j_4  \alpha  \{ \cos ( \theta + \varphi_1 ) + \cos ( \theta - \varphi_2 )  \}.
\end{eqnarray}
The classical exchange energy shows a minimum at $\theta = 114^{\circ}$ when $J_4/J_2 \equiv j_4 = 0.53$ and 
$\theta = 119^{\circ}$ when $j_4 = 0.58$. 
Consequently, the coplanar spiral magnetic structure in SrMn$_3$P$_4$O$_{14}$ can be explained by 
the frustration between the NN and NNN exchange interactions in the chains. 
The value of $k_y$ at 1.8 K is close to 1/3. 
We cannot determine whether the magnetic structure above $T_{\rm N2}$  
is commensurate or not. 
In the frustrated trimerized chain model, 
it is not meaningful whether commensurate or not. 
The following problems should be solved in further investigations. 
$E_{\rm ex}$ is unchanged when we exchange $\varphi_1$ and $- \varphi_2$. 
$E_{\rm ex}$ can be a minimum when  $\varphi_1 = - \varphi_2$. 
Therefore, we cannot deduce the experimental values of $\varphi_1$ and $\varphi_2$ 
from this model. 
In addition, this model cannot explain the reason why the coplanar magnetic structure appears. 
We should consider anisotropy. 

At present, the value of $J_2$ is not determined. 
If the $J_2$ value is small, dipole-dipole (dd) interactions between magnetic moments 
in the $J_2$ or $J_4$ bonds may affect the $\theta$ value. 
Therefore, we consider the following Hamiltonian of the dipole-dipole interactions in the $J_2$ and $J_4$ bonds. 
\begin{eqnarray}
{\cal H}_{\rm dd} &=& \frac{\mu_0 (g \mu_{\rm B})^2}{4 \pi} \Sigma_i
[ \nonumber \\ & & +
\frac{1}{R_2^3} 
\{
{\bf S}_{21,i} \cdot {\bf S}_{22,i-1} -3 \frac{({\bf S}_{21,i} \cdot {\bf R}_2) ({\bf S}_{22,i-1}  \cdot {\bf R}_2)}{R_2^2}
\}
\nonumber \\ & & +
\frac{1}{R_4^3} 
\{
{\bf S}_{1,i} \cdot {\bf S}_{21,i+1} -3 \frac{({\bf S}_{1,i} \cdot {\bf R}_4) ({\bf S}_{21,i+1} \cdot {\bf R}_4)}{R_4^2}
\}
\nonumber \\ & & +
\frac{1}{R_4^3} 
\{
{\bf S}_{1,i} \cdot {\bf S}_{22,i-1} -3 \frac{({\bf S}_{1,i} \cdot {\bf R}_4) ({\bf S}_{22,i-1} \cdot {\bf R}_4)}{R_4^2}
\}
\nonumber \\ & & ].
\end{eqnarray}
The vectors ${\bf R}_2$ and ${\bf R}_4$ are defined in Fig. 1. 
We calculated the classical energy of the dd interactions. 
The first term in the parentheses \{ \} can be renormalized in the classical exchange energy. 
A part of the second term can be also renormalized, while 
the other part of the second term cannot be renormalized. 
In incommensurate magnetic structures, 
the angle between a magnetic moment and the $a$ axis ($\phi$) can have any value in equal probability. 
We obtained the calculated results that the classical dd energy remaining after the $\phi$ average 
can be renormalized in the classical exchange energy. 
The $\theta$ dependent part of the classical total energy per trimer is expressed as follows. 
\begin{eqnarray}
E'( \theta ) &=& M_2^2 \cos ( \theta + \varphi_1 - \varphi_2 ) 
\{ \frac{J_2}{(g \mu_{\rm B})^2} + \epsilon_{{\rm dd}, 2} \} \nonumber \\
& & + M_1 M_2 \{ \cos ( \theta + \varphi_1 ) + \cos ( \theta - \varphi_2 ) \} \nonumber \\
& & \{ \frac{J_4}{(g \mu_{\rm B})^2} + \epsilon_{{\rm dd}, 4} \},
\end{eqnarray}
\begin{equation}
\epsilon_{{\rm dd}, j} = \frac{\mu_0}{4 \pi R_j^3} ( 1 - \frac{3}{2} \frac{R_{jx}^2 + R_{jz}^2}{R_j^2} ) \ \ \ \ \
(j = 2 \ {\rm or} \ 4).
\end{equation}

As was described, 
the classical exchange energy $E'( \theta )_{\rm ex}$ shows a minimum at $\theta = 114^{\circ}$ 
when $j_4 = 0.53$ and when we use the experimental values at 1.5 K for the other parameters. 
Similarly, 
the classical total energy $E'( \theta )$ shows a minimum when 
\begin{equation}
\{ \frac{J_4}{(g \mu_{\rm B})^2} + \epsilon_{{\rm dd}, 4} \} / \{ \frac{J_2}{(g \mu_{\rm B})^2} + \epsilon_{{\rm dd}, 2} \}
= 0.53.
\end{equation}
Figure 7 shows $j_4$ vs. $J_2 /k$ determined from Eq. (7). 
The values of $\epsilon_{{\rm dd}, 2}$ and $\epsilon_{{\rm dd}, 4}$ were calculated as 
$-6.66 \times 10^{20}$ and $4.69 \times 10^{20}$ T$^2$/J, respectively. 
For example, when $J_2 /k = 1$ K and $g = 1.98$,\cite{Hase09} 
the value of $\frac{J_2}{(g \mu_{\rm B})^2}$ is $4.98 \times 10^{22}$ T$^2$/J and 
is much larger than the values of $\epsilon_{{\rm dd}, 2}$ and $\epsilon_{{\rm dd}, 4}$. 
Therefore, the dd interactions do not play an important role for the appearance of the spiral magnetic structure 
when $J_2 /k = 1$ K. 
The $j_4$ value is 0.51 when $J_2 /k = 1$ K (Fig. 7).
The change of $j_4$ caused by the introduction of the dd interactions is small. 
When $J_2 /k = 0.1$ K, $j_4 = 0.33$. 
Influence of the dd interactions cannot be negligible in this case. 

\begin{figure}
\begin{center}
\includegraphics[width=8cm]{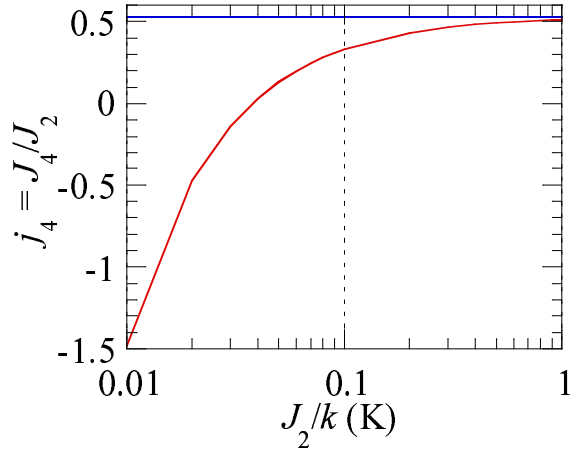}
\caption{
(Color online)
$J_4/J_2 \equiv j_4$ vs. $J_2 /k$ calculated from Eq. (7) (red line). 
The horizontal blue line shows the value (0.53) without the consideration of the dd interactions.   
}
\end{center}
\end{figure}

We briefly comment on the origin of the phase transition at $T_{\rm N2} = 1.75(5)$ K. 
As was shown in Eq. (5), 
the classical dd energy is renormalized in the classical exchange energy. 
Therefore, the dd interactions cannot be a driving force of the phase transition. 
We speculate that the value of $j_4$ is changed at $T_{\rm N2}$ 
by a structural phase transition which we have not detected. 

We describe another origin of spiral magnetic structures. 
Competition between symmetric exchange and Dzyaloshinskii-Moriya (DM) interactions 
can produce spiral magnetic structures. 
Examples of materials are Ba$_2$CuGe$_2$O$_7$,\cite{Zheludev97} CsCuCl$_3$,\cite{Jacobs98} and 
CuB$_2$O$_4$.\cite{Roessli01} 
A uniform component of DM vectors is necessary. 
In SrMn$_3$P$_4$O$_{14}$, the $J_1$ interaction is dominant. 
The DM interaction between Mn1 and Mn2 spins in the $J_1$ bond may exist because of the symmetry. 
The Mn1 site is an inversion center. 
The DM vector between Mn1 and Mn21 spins is anti-parallel to the DM vector between Mn1 and Mn22 spins, 
{\it i.e.,} the DM vectors have no uniform component. 
Accordingly, the DM interaction in the $J_1$ bond 
cannot be the origin of the spiral magnetic structure in SrMn$_3$P$_4$O$_{14}$. 
The $J_2$ interaction is probably the second dominant. 
The DM interaction between Mn21 and Mn22 spins in the $J_2$ bond cannot exist because of the symmetry. 

\section{Conclusion}

We performed neutron powder diffraction experiments on 
the spin-5/2 antiferromagnetic trimerized chain substance SrMn$_3$P$_4$O$_{14}$.  
The coplanar spiral magnetic structure appears below $T_{\rm N1} = 2.2(1)$ K.  
The propagation vector is parallel to the $y$ axis with a typical $y$ component of 0.3. 
The magnetic moments lie in the $ac$ plane in the monoclinic crystal structure. 
The magnetic diffuse scattering is apparent at low temperatures and remains even at 1.5 K.
The shape of the diffuse scattering resembles one- or two-dimensional magnetic Bragg scattering 
with a cutoff at low Q and long tail at high Q (scattering vector magnitude).  
Values of several magnetic structure parameters change rapidly at $T_{\rm N2} = 1.75(5)$ K, 
indicating another phase transition, 
although the magnetic structures above and below $T_{\rm N2}$ are the qualitatively same.  
The spiral magnetic structure can be explained by 
the frustration between the nearest-neighbor and next-nearest-neighbor exchange interactions 
in the trimerized chains. 

\begin{acknowledgments}

This work was partially supported by grants from NIMS.
The neutron powder diffraction experiments were conducted
at SINQ, PSI Villigen, Switzerland. 
We are grateful 
to T. Masuda, T. Arima, M. Kohno, and H. Sakurai for invaluable discussion. 

\end{acknowledgments}


\end{document}